%% file: main.tex
\documentclass[conference,english]{IEEEtran}
\IEEEoverridecommandlockouts

\input{modules/packages}

\input{modules/macros}
\input{modules/abbreviations}

\begin{document}

\title{Distributed Deep Joint Source-Channel Coding over a Multiple Access Channel\\
\thanks{The present work has received funding from the European Union’s Horizon 2020 Marie Skłodowska Curie Innovative Training Network Greenedge (GA. No. 953775) and from CHIST-ERA project SONATA (CHIST-ERA-20-SICT-004) funded by EPSRC-EP/W035960/1. For the purpose of open access, the authors have applied a Creative Commons Attribution (CC BY) license to any Author Accepted Manuscript version arising from this submission.}
}


\author{\IEEEauthorblockN{Selim F. Yilmaz\IEEEauthorrefmark{2}, Can Karamanl{\i}\IEEEauthorrefmark{2}, Deniz Gündüz\IEEEauthorrefmark{2}}
\IEEEauthorblockA{\IEEEauthorrefmark{2}Department of Electrical and Electronic Engineering, Imperial College London, UK \\ \{s.yilmaz21, c.karamanli17, d.gunduz\}@imperial.ac.uk}}

\maketitle

\input{sections/abstract}

\input{sections/introduction}

\input{sections/related_works}

\input{sections/system_model}

\input{sections/methodology}

\input{sections/numerical_results}

\input{sections/conclusion}

\bibliographystyle{IEEEtran}  
\bibliography{main}


\end{document}

%% file: modules/packages.tex
\usepackage{cite}
 \usepackage{amsmath,amssymb,amsfonts,amsthm}
\usepackage[pdftex]{graphicx}
\usepackage{textcomp}
\usepackage{xcolor}
\usepackage{booktabs}
\usepackage{multirow}
\usepackage{multicol}
\usepackage{graphicx}
\usepackage[caption=false]{subfig}
\usepackage[T1]{fontenc}
\usepackage[utf8]{inputenc}
\usepackage{babel}
\usepackage{mathtools}
\usepackage{algorithm}
\usepackage{tikz}
\usepackage{pgfplots}
\usepackage{glossaries-extra}
\usepackage{siunitx}
\usepackage{algpseudocodex}
\usepackage{svg}
\usepackage{etoolbox}
\usepackage{balance}
\usepackage{microtype}
\usepackage{mathtools}
\usepackage[hidelinks]{hyperref}
\usepackage[capitalise]{cleveref}
\usepackage{xcolor, soul}
\sethlcolor{red}

\setabbreviationstyle[acronym]{long-short}

\svgsetup{inkscapelatex=false}

\pgfplotsset{compat=newest,
every axis/.style={
    grid=major,
    xlabel near ticks,
    ylabel near ticks,
    legend pos=south east,
    legend style={font=\footnotesize},
    every x tick label/.append style = {font=\footnotesize},
    every y tick label/.append style = {font=\footnotesize},
    cycle list/Set1-6,
    cycle multiindex* list={
    mark list\nextlist
    Set1-6\nextlist
    linestyles\nextlist
    },
    every axis plot/.append style={mark size=1pt, thick},
    enlarge y limits=0.025,
    enlarge x limits=0
}
}
\graphicspath{{figures/}}

\usepgfplotslibrary{colormaps}
\usepgfplotslibrary{colorbrewer}
\usepgfplotslibrary{external}

\makeatletter
\newcommand\fs@betterruled{%
  \def\@fs@cfont{\bfseries}\let\@fs@capt\floatc@ruled
  \def\@fs@pre{\vspace*{5pt}\hrule height.8pt depth0pt \kern2pt}%
  \def\@fs@post{\kern2pt\hrule\relax}%
  \def\@fs@mid{\kern2pt\hrule\kern2pt}%
  \let\@fs@iftopcapt\iftrue}
\floatstyle{betterruled}
\restylefloat{algorithm}
\makeatother

\usepackage[none]{hyphenat}

%% file: modules/macros.tex
\renewcommand{\vec}[1]{\mathbf{#1}}
\newcommand{\vecs}[1]{\boldsymbol{#1}}

\newcommand{\nv}{\vec{n}}

\newcommand{\rv}{\vec{r}}

\newcommand{\xv}{\vec{x}}
\newcommand{\yv}{\vec{y}}
\newcommand{\zv}{\vec{z}}

\newcommand{\Thetav}{\vecs{\Theta}}
\newcommand{\Phiv}{\vecs{\Phi}}

\newcommand{\Cc}{{\cal C}}
\newcommand{\Dc}{{\cal D}}

\newcommand{\Lc}{{\cal L}}

\newcommand{\Nc}{{\cal N}}

\newcommand{\Pc}{{\cal P}}

\newcommand{\CC}{\mathbb{C}}

\newcommand{\NN}{\mathbb{N}}
\newcommand{\RR}{\mathbb{R}}
\newcommand{\II}{\mathbb{I}}

\newcommand{\LB}{\left(}
\newcommand{\RB}{\right)}

\newcommand{\LSB}{\left[}
\newcommand{\RSB}{\right]}

\newcommand\norm[1]{\left\lVert#1\right\rVert}

\newcommand{\card}[1]{\vert{#1}\vert}
\newcommand{\logn}[2]{\mathop{\mathrm{log}_{#1} \LB #2\RB}}

\newcommand{\Pavg}{P_\mathrm{avg}}

\newcommand{\Dctrain}{\Dc_\mathrm{train}}
\newcommand{\Dcval}{\Dc_\mathrm{val}}

\newcommand{\removed}[1]{}

%% file: modules/abbreviations.tex
\newacronym{ACM}{ACM}{adaptive coding and modulation}
\newacronym{ADC}{ADC}{analog-to-digital conversion}
\newacronym{AGC}{AGC}{automatic gain control}
\newacronym{AWGN}{AWGN}{additive white Gaussian noise}
\newacronym{BER}{BER}{bit error rate}
\newacronym{BLER}{BLER}{block error rate}
\newacronym{BP}{BP}{backpropagation}
\newacronym{BPTT}{BPTT}{backpropagation through time}
\newacronym{CE}{CE}{cross-entropy}
\newacronym{CFO}{CFO}{carrier frequency offset}
\newacronym{CSI}{CSI}{channel state information}
\newacronym{DAC}{DAC}{digital-to-analog conversion}
\newacronym{DL}{DL}{deep learning}
\newacronym{DFT}{DFT}{discrete Fourier transform}
\newacronym{FFT}{FFT}{fast Fourier transform}
\newacronym{GAN}{GAN}{generative adversarial network}
\newacronym{GRU}{GRU}{gated recurrent unit}
\newacronym{iid}{i.i.d.\@}{independent and identically distributed}
\newacronym{IFFT}{IFFT}{inverse fast Fourier transform}
\newacronym{KL}{KL}{Kullback-Leibler}
\newacronym{LSTM}{LSTM}{long short-term memory}
\newacronym{MDP}{MDP}{Markov decision process}
\newacronym{ML}{ML}{machine learning}
\newacronym{MLP}{MLP}{multilayer perceptron}
\newacronym{MIMO}{MIMO}{multiple-input multiple-output}
\newacronym{MSE}{MSE}{mean squared error}
\newacronym{NN}{NN}{neural network}
\newacronym{DNN}{DNN}{deep neural network}
\newacronym{OFDM}{OFDM}{orthogonal frequency-division multiplexing}
\newacronym{pdf}{pdf}{probability density function}
\newacronym{pmf}{pmf}{probability mass function}
\newacronym{PSNR}{PSNR}{peak signal to noise ratio}
\newacronym{RBF}{RBF}{Rayleigh block-fading}
\newacronym{ReLU}{ReLU}{rectified linear unit}
\newacronym{RL}{RL}{reinforcement learning}
\newacronym{RNN}{RNN}{recurrent neural network}
\newacronym{SFO}{SFO}{sampling frequency offset}
\newacronym{SNR}{SNR}{signal-to-noise ratio}
\newacronym{SINR}{SINR}{signal-to-interference-plus-noise ratio}
\newacronym{SGD}{SGD}{stochastic gradient descent}
\newacronym{wrt}{w.r.t.\@}{with respect to}

\newacronym{OAC}{OAC}{over-the-air computation}
\newacronym{MAC}{MAC}{multiple access channel}
\newacronym{SIC}{SIC}{successive interference cancellation}
\newacronym{TDMA}{TDMA}{time division multiple access}
\newacronym{NOMA}{NOMA}{non-orthogonal multiple access}
\newacronym{CL}{CL}{curriculum learning}
\newacronym{JSCC}{JSCC}{joint source-channel coding}
\newacronym{DeepJSCC}{DeepJSCC}{deep joint source-channel coding}
\newacronym{MTL}{MTL}{multi-task learning}
\newacronym{MIL}{MIL}{multi-instance learning}
\newacronym{DML}{DML}{deep metric learning}
\newacronym{IoT}{IoT}{Internet of Things}
\newacronym{SSIM}{SSIM}{structural similarity index measure}
\newacronym{MS-SSIM}{MS-SSIM}{multi-scale \gls{SSIM}}

%% file: sections/abstract.tex
\begin{abstract}
We consider distributed image transmission over a noisy \gls{MAC} using \gls{DeepJSCC}. 
It is known that Shannon's separation theorem holds when transmitting independent sources over a \gls{MAC} in the asymptotic infinite block length regime. However, we are interested in the practical finite block length regime, in which case separate source and channel coding is known to be suboptimal. We introduce a novel joint image compression and transmission scheme, where the devices send their compressed image representations in a non-orthogonal manner. While \gls{NOMA} is known to achieve the capacity region, to the best of our knowledge, non-orthogonal \gls{JSCC} scheme for practical systems has not been studied before. Through extensive experiments, we show significant improvements in terms of the quality of the reconstructed images compared to orthogonal transmission employing current \gls{DeepJSCC} approaches particularly for low bandwidth ratios. We publicly share source code to facilitate further research and reproducibility.
\end{abstract}

\begin{IEEEkeywords}
	Joint source-channel coding, non-orthogonal multiple access, wireless image transmission, deep learning.
\end{IEEEkeywords}

\glsresetall

%% file: sections/introduction.tex
\section{Introduction}
\begin{figure}[t!]
\centering
\subfloat{\includegraphics[width=\columnwidth]{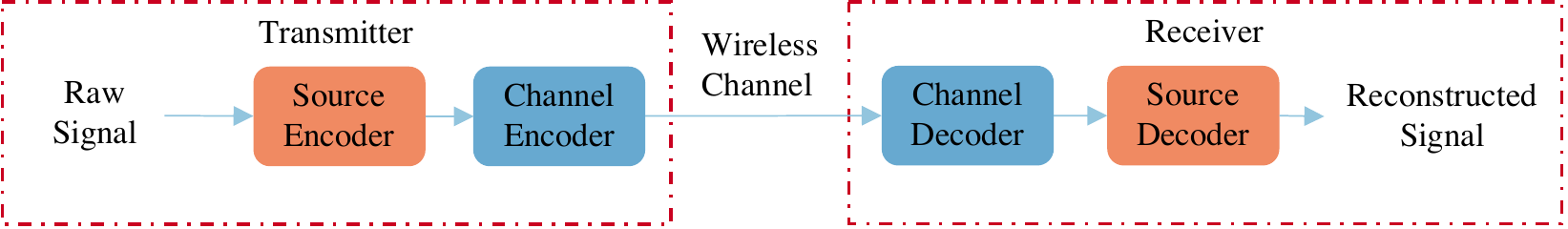}} \\
\subfloat{\includegraphics[width=0.75\columnwidth]{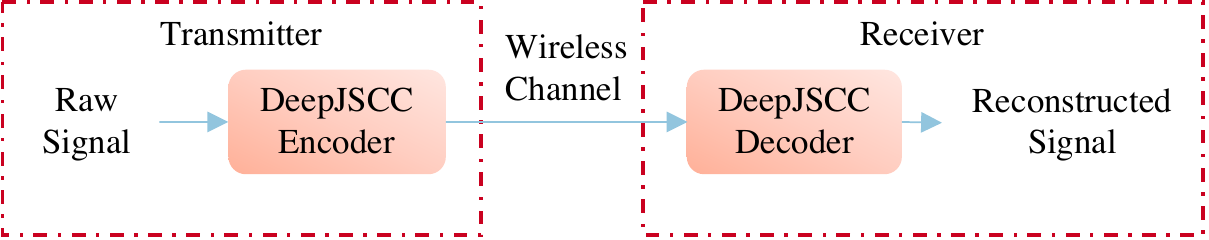}}
\caption{Separation and DeepJSCC-based data transmission schemes}
\label{fig:data_transmission}
\end{figure}

Conventional wireless communication systems consist of two different stages, called source coding and channel coding. Source coding compresses a signal by removing inherent redundancies. Afterwards, channel coding introduces structured redundancy to improve reliability against channel's corrupting effects, such as noise and fading. The receiver has to revert these steps by employing a channel decoder and a source decoder, respectively. For instance, wireless image transmission systems employ JPEG or BPG compression to reduce the required communication resources at the expense of reduced reconstructed image quality. The system then employs channel coding, such as LDPC, turbo codes, or polar coding for reliable transmission over a noisy channel. \Cref{fig:data_transmission} illustrates the conventional separation-based system design for wireless data transmission.

Shannon proved that such a separate design is without loss of optimality when the source and channel block lengths are infinite~\cite{shannon1948mathematical}. However, optimality of separation fails in the practical finite block length regime, or in multi-user networks in general~\cite{gunduz2009source}. Moreover, suboptimality gap of separation increases as the block length gets shorter or the \gls{SNR} diminishes, which correspond to the operation of many delay-sensitive applications such as \gls{IoT} or autonomous driving.

Although researchers have been investigating \gls{JSCC} schemes for many decades~\cite{ramchandran1993multiresolution,zhai2005joint,bursalioglu2013joint}, these schemes did not find application in practical systems due to their high complexity, and poor performance with practical sources and channel distributions. Research on \gls{JSCC} has gained renewed interest recently with the adoption of \glspl{DNN} to implement \gls{JSCC}~\cite{bourtsoulatze2019deep}. This data-driven approach is based on modeling the end-to-end communication system as an autoencoder architecture. Numerous follow-up studies have extended \gls{DeepJSCC} in different directions. In \cite{kurka2020deepjscc}, the authors demonstrate that \gls{DeepJSCC} can exploit feedback to improve the end-to-end reconstruction quality. In \cite{burth2020joint}, \gls{DeepJSCC} architecture is modified by increasing the filter size to improve the performance. 
In~\cite{kurka2021bandwidth}, it is shown that \gls{DeepJSCC} can adopt to varying channel bandwidth with almost no loss in performance. In addition to very promising results achieved for a particular channel quality, it is worth noting that \gls{DeepJSCC} avoids the \textit{cliff effect} suffered by separate source and channel coding and achieves \textit{graceful degradation} with the channel quality; that is the image is still decodable even when the channel quality falls below the target \gls{SNR} the system is designed for, albeit with lower quality. Separation-based schemes, on the other hand, completely fail as the channel code cannot be decoded below a certain \gls{SNR} threshold.


In this paper, we consider transmission of distinct images over a \gls{MAC}. It is known that separation theorem applies to 
\glspl{MAC} with independent sources; however, the design of practical \gls{JSCC} schemes for finite block lengths and real information sources remain as an open challenge. A straightforward approach is to employ \gls{TDMA}, where each user employs the same trained \gls{DeepJSCC} network to transmit using half of the channel bandwidth. However, our goal in this work is to develop a \gls{DeepJSCC} framework that can exploit non-orthogonal communications. A related NOMA-aided JSCC scheme is considered for collaborative inference over a MAC in~\cite{lo2023collaborative}, where the goal is to retrieve a unique identity rather than reconstructing distinct images.

Our main contributions are summarized as follows:
\begin{enumerate}
    \item To the best of our knowledge, we introduce the first multi-user \gls{DeepJSCC}-based image transmission method, exploiting \gls{NOMA}.
    \item We introduce a novel Siamese network-based architecture with device embeddings that allows us to use the same set of parameters for both devices, which will be instrumental in scaling our solution to larger networks.
    \item We introduce a training data subsampling methodology, and employ \gls{CL}-based training to further improve the performance.
    \item Through an extensive set of experiments, we show that our superposition-based method for \gls{NOMA} outperforms traditional DeepJSCC with time division for all evaluated \gls{SNR} conditions. We also show that our method is fair; that is, no significant difference is observed between the performances of the two users.
    \item To facilitate further research and reproducibility, we provide the source code of our framework and simulations on \href{https://github.com/ipc-lab/deepjscc-noma}{github.com/ipc-lab/deepjscc-noma}.
\end{enumerate}


%% file: sections/related_works.tex
\section{Related Works}
\label{sec:related_works}
In this section, we review the works that we construct our methods upon.
\subsection{Non-Orthogonal Multiple Access (NOMA)}

\begin{figure}
    \centering
    \includegraphics[width=0.75\columnwidth]{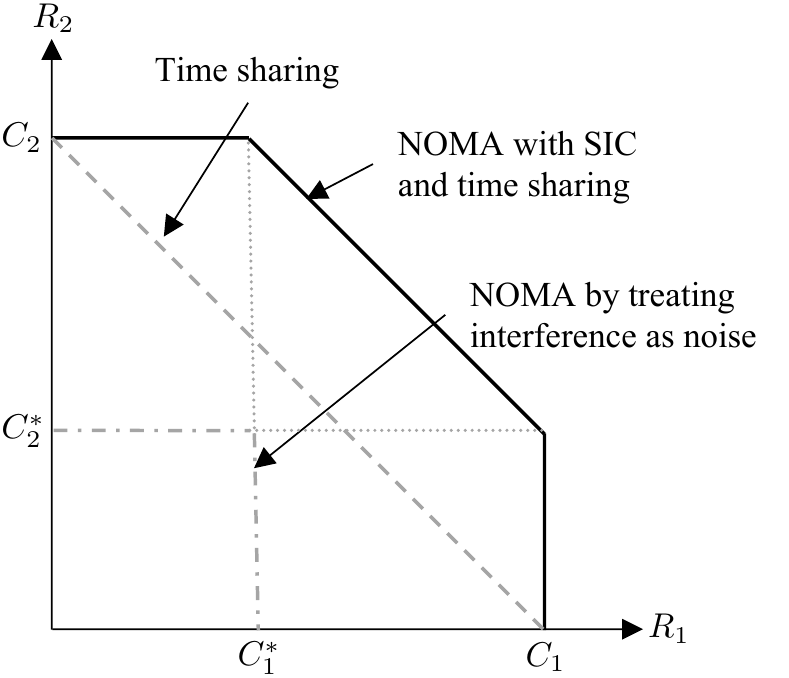}
    \caption{Capacity region of a two-user \gls{MAC} compared with the rate pairs achievable by \gls{TDMA}~\cite{goldsmith2005wireless}.}
    \label{fig:rates}
\end{figure}

\gls{NOMA} is essential in achieving the capacity region of a \gls{MAC}, whereas orthogonal transmission schemes, such as \gls{TDMA}, are strictly suboptimal. Please see \cref{fig:rates} for an illustration of the capacity region of a \gls{MAC}, achievable by \gls{NOMA}, and the suboptimality of \gls{TDMA}. Although the signals from different users act as interference to each other, either joint decoding~\cite{cover1999elements}, or \gls{SIC} with message splitting~\cite{grant2001rate} allows to achieve higher rates and meet the capacity. Recently, there have also been efforts to employ \glspl{DNN} to implement SIC for \gls{NOMA}~\cite{shlezinger2020deepsic,sim2020deep,van2022deep}. 
On the other hand, in DeepJSCC, input signal are directly mapped to channel inputs without imposing any constellation constraints. The continuous-amplitude nature of the transmitted signals is beneficial in achieving graceful degradation with channel quality; however, it also means that the decoder is an estimator, and will always have some noise in its reconstruction. That is, unlike in digital communication, the decoder cannot recover the transmitted codeword perfectly, which means that perfect interference cancellation is not possible. 



\subsection{Related Learning Paradigms: Multi-Task Learning (MTL), Deep Metric Learning (DML) and Curriculum Learning (CL)}
MTL is a paradigm of learning multiple tasks jointly to improve performance~\cite{caruana1997multitask,zhang2021survey}. In the \gls{MAC}, multiple signals are superposed onto a joint representation, and the decoder reconstructs multiple images. Therefore, it is inherently a MTL problem. In distributed compression~\cite{mital2022neural}, typically different encoding and decoding functions are employed for each of the source signals, which increases the complexity and inhibits scalability. MTL methods often modify feature space or share parameters to improve performance for both tasks~\cite{zhang2021survey}. Inspired from these tricks in the MTL literature, we introduce a novel input augmentation embedding for better separability and employ a Siamese neural network-based architecture to reduce the number of parameters~\cite{koch2015siamese}.

Similarly to DML, our method projects multiple instances into a lower-dimensional latent space with the same network. In DML, typically a subsample of pairs are used for training, and the choice of training pairs significantly affects performance~\cite{hoffer2015deep}. Therefore, we design a systematic and efficient training methodology instead of using all possible training pairs, which is infeasible.

\Gls{CL} is a progressive training paradigm, where network is first trained on easy tasks and then adapted to the main task of interest~\cite{bengio2009curriculum}. \Gls{CL} has been previously leveraged in~\cite{shao2022attentioncode} by starting training from higher \glspl{SNR} for progressively adapting to lower \glspl{SNR}. In our problem, since signals are superposed, they interfere with each other, which affects the training performance. Therefore, we further improve our method by training the network first on non-interfering signals, and then fine-tune it on the actual task.

%% file: sections/system_model.tex
\textbf{Notation:} Unless stated otherwise; boldface lowercase letters denote vectors (e.g., $\vec{p}$), boldface uppercase letters denote matrices (e.g., $\vec{P}$), non-boldface letters denote scalars (e.g., $p$ or $P$), and uppercase calligraphic letters denote sets (e.g., $\Pc$). $\RR$, $\NN$, $\CC$ denote the set of real, natural and complex numbers, respectively. $\card{\Pc}$ denotes the cardinality of set $\Pc$. We define $[n]\triangleq\{1,2,\cdots,n\}$, where $n\in\NN^+$, and $\II \triangleq [255]$.

\section{System Model}
\label{sec:system_model}
We consider the distributed wireless image transmission problem over a \gls{MAC} in an uplink setting with two transmitters and a single receiver. Specifically, consider \gls{AWGN} channel with noise variance $\sigma^2$. Transmitter $i$ maps an input image $\xv_i \in \II^{C_\mathrm{in} \times W \times H}$, where $W$ and $H$ denote the width and height of the image, while $C_\mathrm{in}$ represents the R, G and B channels for colored images, with a non-linear encoding function $E_{\Thetav_i, \sigma}:\II^{C_\mathrm{in} \times W \times H} \rightarrow \CC^{k}$ parameterized by $\Thetav_i$ into a complex-valued latent vector $\zv_i=E_{\Thetav_i,\sigma} (\xv_i)$, where $k$ corresponds to the available channel bandwidth.

We enforce average transmission power constraint on both transmitters:
\begin{align}
\label{eq:power_constraint}
\frac{1}{k} \norm{\zv_i}_2^2 \leq \Pavg, \,\, i \in \{1,2\}.
\end{align}

The receiver receives the summation of latent vectors as:
\begin{align*}
\yv = \zv_1 + \zv_2 + \nv,
\end{align*}
where $\nv \in \CC^{k}$ is \gls{iid} complex Gaussian noise term with variance $\sigma^2$, i.e., $\nv \sim \Cc\Nc(\vec{0}, \sigma^2 \vec{I}_{k})$. We consider $\sigma$ is known at the transmitters and the receiver.

Then, a non-linear decoding function  $D_{\Phiv,\sigma}:\CC^{k} \rightarrow \II^{2 \times C_\mathrm{in} \times W \times H}$ at the receiver, parameterized by $\Phiv$, reconstructs both images that are aggregated in the common representation $\yv$, i.e.,
\begin{align*}
    \begin{bmatrix}
           \hat{\vec{x}}_1\\
           \hat{\vec{x}}_2
         \end{bmatrix} = D_{\Phi,\sigma} (\yv).
\end{align*}

The {\it bandwidth ratio} $\rho$ characterizes the available channel resources, and is defined as:
\begin{align*}
    \rho = \frac{k}{ C_\mathrm{in} W H} \,\, \si{channel\,symbols \per pixel}.
\end{align*}
We also define the $\mathrm{SNR}$ at time $t$ as:
\begin{align}
    \label{eq:snr}
    \mathrm{SNR} = 10 \logn{10}{\frac{\Pavg}{\sigma^2}} \,\, \si{\decibel}.
\end{align}

Our objective is to maximize the average \gls{PSNR}, on an unseen target dataset under given channel \gls{SNR} and the power constraint in~\eqref{eq:power_constraint}, which is defined as:
\begin{align}
\label{eq:psnr}
\mathrm{PSNR} = 10 \logn{10}{\frac{A^2}{
    \frac{1}{C_\mathrm{in}HW}
    \norm{\xv_i - \hat{\xv}_i}_2^2 }} \,\, \si{\decibel},
\end{align}
where $A$ is the maximum possible input value, e.g., $A=255$ for images with 8-bit per channel as in our case.  To achieve this goal, in the next section, we will introduce our framework to implement joint training of the neural network based decoder $D_{\Phiv,\sigma}$ and encoders $E_{\Thetav_i,\sigma}, \,\, i \in \{1,2\}$.

%% file: sections/methodology.tex
\section{Methodology}
\label{sec:methodology}
In this section, we describe our novel methodology to exploit \gls{NOMA} for efficient image transmission using \gls{DeepJSCC}. \Cref{alg:overview} summarizes the training methodology of the introduced method, called DeepJSCC-NOMA.

\begin{figure}[!t]
    \centering
    \includegraphics[width=\columnwidth]{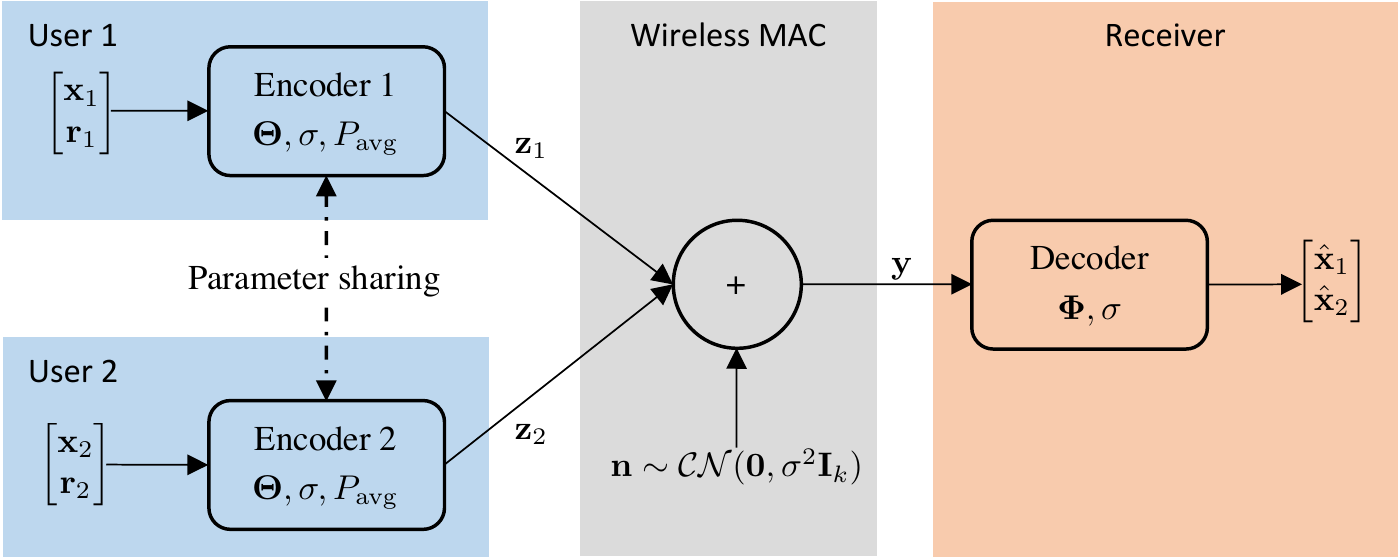}
    \caption{Overall architecture of our DeepJSCC-NOMA method.}
    \label{fig:architecture}
\end{figure}

\subsection{DeepJSCC-NOMA Architecture}
Following the practice of previous works~\cite{bourtsoulatze2019deep,burth2020joint}, we construct our network via an autoencoder-based architecture. We employ the encoder and decoder architectures in~\cite{tung2022deepjscc}, removing the quantization part as our method allows continuous channel inputs. It has symmetric encoder and decoder networks with two downsampling/upsampling layers. It utilizes residual connections and the attention mechanism introduced in~\cite{cheng2020learned}. It also leverages \gls{SNR} adaptivity using the modules proposed in~\cite{xu2021wireless}, named the attention feature (AF) module. This module allows using the same model for test on channels with different \glspl{SNR} without significant degradation in performance. It allows the network to learn parameters for different \glspl{SNR} by having the \gls{SNR} as an input feature, and training with randomly sampled \glspl{SNR}. This way, it allows using the same model on channels with different \gls{SNR} values without significant degradation in performance.


Note that the architecture in~\cite{tung2022deepjscc} is for point-to-point \gls{DeepJSCC}. Distributed compression architectures such as~\cite{mital2022neural} employ encoders with different parameters for each device. However, this makes training more demanding due to the large number of parameters. Instead, we consider a Siamese neural network architecture~\cite{koch2015siamese}, in which the two encoders at the transmitters share their parameters, i.e., $\Thetav_1=\Thetav_2=\Thetav$. This significantly reduces the complexity thanks to our novel device embedding-based input augmentation method. \Cref{fig:architecture} summarizes our architecture for DeepJSCC-NOMA.

We employ a power normalization layer at the end of each encoder to satisfy the average power constraint in~\eqref{eq:power_constraint}, i.e., we normalize the transmitted signal by multiplying $\zv_i$ with $\sqrt{k \Pavg /\norm{\zv_i}_2^2}$.

\input{figures/alg_overview}

We jointly train the whole network using the training data pairs $\Dctrain$ using the loss function $\Lc$:
\begin{align*}
    \Lc \LB \Thetav, \Phiv \RB = \sum_{\substack{(\xv_1, \xv_2) \\ \in \Dctrain}} \mathrm{MSE} \LB \xv_1, \hat{\xv}_1 \RB + \mathrm{MSE} \LB \xv_2, \hat{\xv}_2 \RB,
\end{align*}
where $\mathrm{MSE}(\xv,\hat{\xv}) \triangleq \frac{1}{m} \norm{\xv - \hat{\xv}}_2^2$ and $m$ is the total number of elements in $\xv$, which is given by $m=C_\mathrm{in} W H$. We note that the introduced method is unsupervised as it does not rely on any costly human labeling, and raw images and the channel model are enough to train our method.

Our method takes multiple instances as input and aims to decode two images, i.e., $\xv_1$ and $\xv_2$, simultaneously. Therefore, our task fits into the MTL~\cite{zhang2021survey} framework. One common method in MTL is to share parameters to model commonalities between tasks. In our problem, the task of image transmission for both encoders is exactly equivalent, so it is natural to share parameters of the encoders. However, the receiver also needs to distinguish between the images transmitted by the two encoders during the joint decoding phase. This is why we employ input augmentation, which is explained in the next section.

\subsection{Input Augmentation via Novel Device Embeddings}
\label{sec:input_augmentation}

Due to the superposition nature of the \gls{MAC}, even if the receiver can recover both images with low distortion, it is challenging in the case of NOMA to distinguish which image belongs to which transmitter. We also want a scalable architecture; hence, we would like to employ the same encoder architecture for both transmitters. Note that an encoder that takes both images as input is not possible since we assume no communication between the devices during  the transmission phase.

Without significantly increasing the number of parameters and without changing the \gls{DeepJSCC} architecture that is known to perform well, we introduce an input augmentation method to be able to improve separability of signals in the aggregate representation domain. Hence, we employ device embeddings, which are unique trainable weights for each transmitting device. We initialize the two trainable device embeddings $\rv_i \in 1 \times W \times H$ randomly from $\Nc \LB 0,1 \RB$, $i=1,2$, and concatenate with the input image as follows:
\begin{align*}
    \tilde{\xv}_i \triangleq
    \begin{bmatrix}
    \xv_i\\ \rv_i
    \end{bmatrix}.
\end{align*}
Therefore, the inputs will have $C_\mathrm{in}+1$ dimensions, e.g., 4 dimensions instead of 3 for RGB images. This simple method allows the network to differentiate between different devices. These embeddings are optimized jointly during training and remain constant during test time. For $i = 1,2$, $\tilde{\xv}_i$ is given to encoder $i$ as input, instead of $\xv_i$, to obtain $\zv_i = E_{\Thetav,\sigma} (\tilde{\xv}_i)$ before power normalization.

Instead of separate decoders at the receiver, we only employ image specific parameters at the last layer of the decoder. This significantly reduces the number of parameters and eases the training process while better separating the transmitted signals. The decoder $D_{\Phiv, \sigma}$ outputs $2C_\mathrm{in} \times W \times H$ dimensional image tensor, which is then reshaped into two separate images, i.e., to the shape $2 \times C_\mathrm{in} \times W \times H$. Notice that although device embeddings introduce new trainable  parameters, it only depends on the image size, and does not scale with the number of devices.

Notice that, by minimizing $\Lc$, the network learns to model intra-user distributions $p(\hat{\xv}_1 | \tilde{\xv}_1)$, $p(\hat{\xv}_2 | \tilde{\xv}_2)$ as well as inter-user distributions  $p(\hat{\xv}_2 | \tilde{\xv}_1)$, $p(\hat{\xv}_1 | \tilde{\xv}_2)$. With inter-user distributions, the encoders ease separability of the input instances by training device embeddings $\rv_1$ and $\rv_2$ since the inputs $\xv_1$ and $\xv_2$ are independent and the encoders share all the parameters.

\subsection{Training Data Subsampling Strategy}
\label{sec:subsampling}
We now need to define the training pairs in $\Dctrain$ to compute the loss, $\Lc$. Since we are training the network on pairs $(\xv_1, \xv_2)$, there are $N^2$ possible pair combinations for a given training dataset with $N$ samples. However, it is generally infeasible to use all of these pairs since the number of training instances grows quadratically with N. Therefore, we subsample $T \ll N^2$ pairs from all possible combinations as in DML problems~\cite{hoffer2015deep}. To ensure uniqueness of every pair, we subsample without replacement. We use the same pairs in $\Dctrain$ after shuffling to minimize $\Lc$. We also construct validation pairs $\Dcval$ by splitting the initial validation data into two and use the same validation pairs after every epoch.

\subsection{CL-Based Training Methodology}
\label{sec:curriculum_learning}
In our problem, the received signals are very noisy due to the interference from the other device. We address this problem by first training the network without superposition (which is an easier task) by only using the signal and the \gls{AWGN} noise, i.e., by computing $\hat{\xv}_1$ via $D_{\Phiv,\sigma}(\zv_1 + \nv)$ and  $\hat{\xv}_2$ via $D_{\Phiv,\sigma}(\zv_2 + \nv)$, both with the power constraint $\Pavg$ as in our standard training strategy. We then fine-tune the network with the standard training strategy on superposed signals as described above. This way, we enjoy benefits of the progressive training process via \gls{CL}. Note that we do not change the loss function or any other component of the network in any phase of our \gls{CL} training strategy.


%% file: figures/alg_overview.tex
\begin{algorithm}[t]
    \caption{\strut Training Procedure of DeepJSCC-NOMA}
    \label{alg:overview}
    \begin{algorithmic}
    \State Construct $\Dctrain$ and $\Dcval$
    \Comment{See Section~\ref{sec:subsampling}}
    \State Initialize $\rv_1$ and $\rv_2$ from $\Nc \LB 0,1 \RB$
    \Comment{Device embeddings}
    \Repeat
    \Comment{Iterate through epochs}
        \State Shuffle $\Dctrain$ and set $\Lc=0$
        \ForAll{$(\xv_1,\xv_2) \in \Dctrain$}
            \State Calculate $\sigma$ via~\eqref{eq:snr} for $\mathrm{SNR} \sim \mathrm{Uniform}\LSB 0,20 \RSB$
            \For{$i=1,2$}
            \Comment{Device with index $i$}
                \State $\tilde{\xv}_i = \begin{bmatrix} \xv_i & \rv_i \end{bmatrix}^T$
                \State $\zv_i = E_{\Thetav,\sigma} (\tilde{\xv}_i)$
                \Comment{Encoding}
                \State $\zv_i = \sqrt{k \Pavg/\norm{\zv_i}_2^2} \zv_i$
                \Comment{Power normalization}
            \EndFor
            \LComment{Transmit $\zv_1$ and $\zv_2$ via same channel}
            \State Sample $\nv \sim \Cc\Nc(\vec{0}, \sigma^2 \vec{I}_{k})$
            \State $\yv = \zv_1 + \zv_2 + \nv$
            \Comment{Received signal}
            \State $\begin{bmatrix} \hat{\xv}_1 & \hat{\xv}_2 \end{bmatrix}^T = D_{\Phi,\sigma} (\yv)$
            \Comment{Decoding}
            \State $\Lc = \Lc + \mathrm{MSE} \LB \xv_1, \hat{\xv}_1 \RB + \mathrm{MSE} \LB \xv_2, \hat{\xv}_2 \RB$
            \If{$t \,\, \mathrm{mod} \,\, batch\, size = 0$}
                \LComment{Standard mini-batch training}
                \State Update $\rv_1, \rv_2, \Thetav, \Phiv$ by backpropagating $\Lc$ 
                \State Reset $\Lc$ and gradients to zero
            \EndIf
        \State Compute validation PSNR using $\Dcval$
        \EndFor
    \Until{validation PSNR does not improve for $e$ epochs}
    \end{algorithmic}
\end{algorithm}

%% file: sections/numerical_results.tex
\input{figures/main_comparison}
\section{Numerical Results}
\label{sec:numerical_results}
In this section, we present our experimental setup to demonstrate the performance gains of our method in different scenarios.
\subsection{Dataset}
We employ CIFAR-10 dataset for training and testing, which has training data with \num{50000} instances and test data with \num{10000}. All the images have the shape of $3 \times 32 \times 32$ as RGB channels' dimension, width and height, respectively. We split the training data into \num{45000} training instances and \num{5000} validation instances.



\subsection{Baselines}
We compare our method with classical \gls{DeepJSCC} with time division, named DeepJSCC-TDMA.  We name our method DeepJSCC-NOMA-CL. We compare it with the model without \gls{CL} described in \cref{sec:curriculum_learning}. We also compare it with the case of strong assumption of ideal \gls{SIC} scenario to show the potential of our method, named DeepJSCC-SingleUser. This is also the initial model in \gls{CL}-based strategy described in \cref{sec:curriculum_learning}, which is trained and tested with non-interfering signals. We also note that previous studies already showed the superiority of \gls{DeepJSCC} over classical separation-based transmission methods under power and bandwidth constraints~\cite{bourtsoulatze2019deep}, and in this work, we show superiority of our method over standard \gls{DeepJSCC} employing time division. 

\subsection{Implementation Details}
We have conducted the experiments using Pytorch framework~\cite{paszke2019pytorch}. We use the same hyperparameters and the same architecture for all the methods. We use learning rate \num{1e-4}\removed{, number of filters in middle layers $256$} and batch size \num{64}. We  set the number of filters in the middle layers to $256$ for both encoder and decoder. We use the power constraint $\Pavg=0.5$ for all the methods to make the TDMA-based DeepJSCC model comparable with previous works. We use Adam optimizer to minimize the loss~\cite{kingma2014adam}. We continue training until no improvement is achieved for consecutive $e=10$ epochs. During training and validation, we run the model using different \gls{SNR} values for each instance, uniformly chosen from $\LSB 0,20 \RSB$ \si{\decibel}. We test and report the results for each \gls{SNR} value using the same model. For the proposed method, we use the training data with \num{200000} unique pairs instead of $\num{45000} \times \num{45000}$ pairs using the method described in \cref{sec:subsampling}. We shuffle the training pairs or instances randomly before each epoch. 

\subsection{Comparison with TDMA-based DeepJSCC}


\Cref{fig:main_comparison} demonstrates the performance gains of our method in different \gls{SNR} conditions for $\rho=1/6$ and $\rho=1/3$ compression ratios. For both compression ratios, DeepJSCC-NOMA performs better  than its \gls{TDMA} counterpart for all evaluated \glspl{SNR}. For $\rho=1/3$, DeepJSCC-NOMA-CL achieves \num{0.91} \si{\decibel} (absolute) higher \gls{PSNR} on average compared to DeepJSCC-TDMA. For $\rho=1/6$, our method achieves \num{0.25} \si{\decibel} (absolute) higher \gls{PSNR} on average compared to DeepJSCC-TDMA. 

\Gls{CL}-based training further improves DeepJSCC-NOMA for all evaluated \glspl{SNR} for both compression ratios. We observe that DeepJSCC-NOMA-CL achieves an improvement of 0.32 \si{\decibel} and 0.08 \si{\decibel} with respect to DeepJSCC-NOMA in terms of \gls{PSNR} for $\rho=1/3$ and $\rho=1/6$, respectively. This shows the benefits of the employed \gls{CL} method. DeepJSCC-SingleUser serves as an upper bound although it is a highly optimistic assumption, so we do not expect it to be tight.


\subsection{Analysis of Fairness}
Figure~\ref{fig:fairness_comparison} illustrates the fairness of DeepJSCC-NOMA-CL for $\rho=1/3$. There is no significant difference between the average \gls{PSNR} achieved by the two devices, while both of them are outperforming \gls{TDMA}-based DeepJSCC. A similar observation is also made for $\rho=1/6$.

\input{figures/fairness_comparison}

\subsection{Analysis of the Number of DNN Parameters}
\label{sec:dnn_parameters}
Table~\ref{tab:num_parameters} shows the comparison of our method with the standard point-to-point \gls{DeepJSCC} with time-division. For both the compression rates $\rho=1/6$ and $\rho=1/3$, the amount of increase in the number of parameters is only $\approx 1\%$ thanks to the input augmentation and single decoder tricks. Otherwise, we would need to use different encoder and decoders as explained in Section~\ref{sec:input_augmentation}, which would cause growth in the number of parameters as the number of devices increases.

Notice that the differences between the proposed DeepJSCC-NOMA method and the classical DeepJSCC for TDMA are the followings: (1) number of parameters in the last layer of the encoder and the first layer of the decoder since the available bandwidth is twice in \gls{NOMA}; (2) trainable embedding of our novel input augmentation described in Section~\ref{sec:input_augmentation}; (3) the number of parameters in the last layer of decoder since we are decoding two different images using the same decoder. But, as the table shows, these introduce only a marginal increase in the number of parameters, and hence, the computational and memory complexity.

\input{figures/tab_num_parameters}



%% file: figures/main_comparison.tex
\begin{figure*}[!t]
    \centering
\begin{tikzpicture}[scale=0.75]
\begin{axis}[
title={Bandwidth Ratio $\rho=1/6$},
xlabel={$\mathrm{SNR}_\mathrm{test}$ (\si{\decibel})},
ylabel={PSNR\ (\si{\decibel})},
xmax=15
]
\addplot table[x=snr, y=cifar10_psnr, col sep=comma]{results/nd2_C12_perfectsic.csv};
\addplot table[x=snr, y=cifar10_psnr, col sep=comma]{results/nd2_C12_singlemodel_ft.csv};
\addplot table[x=snr, y=cifar10_psnr, col sep=comma]{results/nd2_C12_singlemodel.csv};
\addplot table[x=snr, y=cifar10_psnr, col sep=comma]{results/nd1_C12.csv};
\legend{DeepJSCC-SingleUser,DeepJSCC-NOMA-CL,DeepJSCC-NOMA,DeepJSCC-TDMA}
\end{axis}
\end{tikzpicture}
\begin{tikzpicture}[scale=0.75]
\begin{axis}[
title={Bandwidth Ratio $\rho=1/3$},
xlabel={$\mathrm{SNR}_\mathrm{test}$ (\si{\decibel})},
ylabel={PSNR\ (\si{\decibel})},
xmax=15
]
\addplot table[x=snr, y=cifar10_psnr, col sep=comma]{results/nd2_C6_perfectsic.csv};
\addplot table[x=snr, y=cifar10_psnr, col sep=comma]{results/nd2_C6_singlemodel_ft.csv};
\addplot table[x=snr, y=cifar10_psnr, col sep=comma]{results/nd2_C6_singlemodel.csv};
\addplot table[x=snr, y=cifar10_psnr, col sep=comma]{results/nd1_C6.csv};
\legend{DeepJSCC-SingleUser,DeepJSCC-NOMA-CL,DeepJSCC-NOMA,DeepJSCC-TDMA}
\end{axis}
\end{tikzpicture}
\caption{Comparison of our NOMA-based method with TDMA-based \gls{DeepJSCC} for bandwidth ratios $\rho=1/6$ and $\rho=1/3$.}
\label{fig:main_comparison}
\end{figure*}
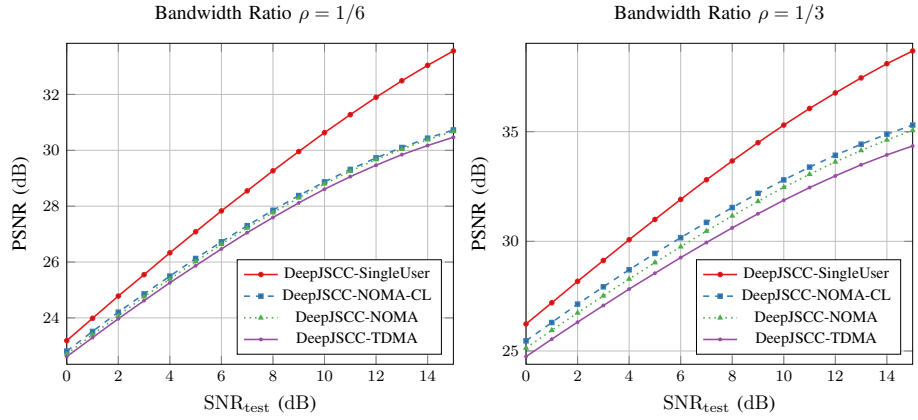

%% file: figures/fairness_comparison.tex
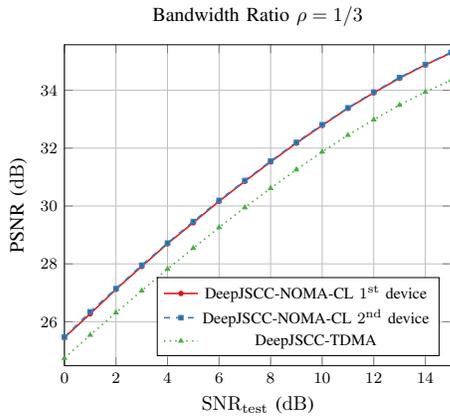
\begin{figure}[!t]
    \centering
\begin{tikzpicture}[scale=0.75]
\begin{axis}[
title={Bandwidth Ratio $\rho=1/3$},
xlabel={$\mathrm{SNR}_\mathrm{test}$ (\si{\decibel})},
ylabel={PSNR\ (\si{\decibel})},
xmax=15
]
\addplot table[x=snr, y=dev0_cifar10_psnr, col sep=comma]{results/fairness_C6_singlemodel_ft.csv};
\addplot table[x=snr, y=dev1_cifar10_psnr, col sep=comma]{results/fairness_C6_singlemodel_ft.csv};
\addplot table[x=snr, y=cifar10_psnr, col sep=comma]{results/nd1_C6.csv};
\legend{DeepJSCC-NOMA-CL $1^\mathrm{st}$ device,DeepJSCC-NOMA-CL $2^\mathrm{nd}$ device, DeepJSCC-TDMA}
\end{axis}
\end{tikzpicture}
    \caption{Fairness analysis between different users of our method for bandwidth ratio $\rho=1/3$.}
    \label{fig:fairness_comparison}
\end{figure}

%% file: figures/tab_num_parameters.tex
\begin{table}[!tbp]
\caption{Number of parameters for the compared methods}
\centering
\begin{tabular}{@{}lrr@{}}
\toprule
Method        & $\rho=1/3$ & $\rho=1/6$ \\ \midrule
DeepJSCC-TDMA & 22.2M      & 22.1M       \\
DeepJSCC-NOMA & 22.4M      & 22.3M       \\
\bottomrule
\end{tabular}
\label{tab:num_parameters}
\end{table}

%% file: sections/conclusion.tex
\section{Conclusion}
\label{sec:conclusion}

We have introduced a novel joint image compression and transmission scheme for the multi-user uplink scenario. The proposed approach exploits \gls{NOMA} and the transmitters employ identical \glspl{DNN}. We have shown that, thanks to the novel input augmentation trick, the receiver is able to recover both images despite the analog transmission of underlying \gls{DeepJSCC} approach, and is able to attribute decoded images to the correct transmitters. We have shown that the proposed DeepJSCC-NOMA scheme achieves superior performance compared to its time-division counterpart.